
%
\input phyzzx
\catcode`@=11
%
%
\newtoks\UT
\newtoks\monthyear
\Pubnum={UT-\the\UT}
\UT={714}
\monthyear={July, 1995}
\def\p@bblock{\begingroup \tabskip=\hsize minus \hsize
    \baselineskip=1.5\ht\strutbox \topspace-2\baselineskip
    \halign to\hsize{\strut ##\hfil\tabskip=0pt\crcr
    \the\Pubnum\cr\the\monthyear\cr
    }\endgroup}
\def\bftitlestyle#1{\par\begingroup \titleparagraphs
    \iftwelv@\fourteenpoint\else\twelvepoint\fi
    \noindent {\bf #1}\par\endgroup}
\def\title#1{\vskip\frontpageskip \bftitlestyle{#1} \vskip\headskip}
%
%
\def\acknowledge{\par\penalty-100\medskip \spacecheck\sectionminspace
    \line{\hfil ACKNOWLEDGEMENTS\hfil}\nobreak\vskip\headskip}
%
%

%
\def\journal#1&#2(#3){\begingroup \let\journal=\dummyj@urnal
    \unskip, \sl #1\unskip~\bf\ignorespaces #2\rm
    (\afterassignment\j@ur \count255=#3) \endgroup\ignorespaces}
\def\andjournal#1&#2(#3){\begingroup \let\journal=\dummyj@urnal
    \sl #1\unskip~\bf\ignorespaces #2\rm
    (\afterassignment\j@ur \count255=#3) \endgroup\ignorespaces}
\def\andvol&#1(#2){\begingroup \let\journal=\dummyj@urnal
    \bf\ignorespaces #1\rm
    (\afterassignment\j@ur \count255=#2) \endgroup\ignorespaces}
\def\IJMP{Int.~J.~Mod.~Phys.}

\def\NP{Nucl.~Phys.}
\def\PL{Phys.~Lett.}
\def\PR{Phys.~Rev.}
\def\PRL{Phys.~Rev.~Lett.}

\catcode`@=12
%


\titlepage

\title{QCD-like Hidden Sector Models
       without the Polonyi Problem}

\author{Izawa {\twelverm Ken-Iti}
\foot{\rm JSPS Research Fellow.}
{\twelverm and Tsutomu} Yanagida}
\address{Department of Physics, University of Tokyo \break
                    Tokyo 113, Japan}

\abstract{QCD-like hidden sector models of
supersymmetry breaking are considered which do not suffer from
a cosmological problem due to the Polonyi field.
Avoidance of a light gluino leads to introduction
of quasi-symmetry
-- symmetry broken explicitly only through gravitational effects.
}

\endpage

\doublespace


\def\a{\alpha}

\def\l{\lambda}

\def\q{\partial}

\def\x{\xi}

\def\L{\Lambda}

\def\j{\journal}
\def\o{\over}


\REF\Nil{For a review, H.P.~Nilles \j \IJMP &A5 (90) 4199.}

\REF\Cou{G.D.~Coughlan, W.~Fischler, E.W.~Kolb, S.~Raby,
         and G.G.~Ross \j \PL &B131 (83) 59.}

\REF\Ban{T.~Banks, D.B.~Kaplan, and A.E.~Nelson
         \j \PR &D49 (94) 779.}

\REF\Aff{I.~Affleck, M.~Dine, and N.~Seiberg \j \NP &B256 (85) 557;
         \nextline N.~Seiberg \j \PL &B318 (93) 469.}

\REF\Wei{S.~Weinberg \j \PRL &48 (82) 1776.}

\REF\Bag{J.~Bagger, E.~Poppitz, and L.~Randall \j \NP &B426 (94) 3.}

\REF\Joi{I.~Joichi and M.~Yamaguchi \j \PL &B342 (95) 111.}

\REF\Cre{E.~Cremmer, S.~Ferrara, L.~Girardello, and A.~Van Proeyen
         \j \PL &B116 (82) 231; \andjournal \NP &B212 (83) 413.}

\REF\Giu{G.F.~Giudice and A.~Masiero \j \PL &B206 (88) 480.}

\REF\Gid{S.~Giddings and A.~Strominger \j \NP &B307 (88) 854.}

\sequentialequations

%
{\caps 1. Introduction}

Nonabelian gauge theories are expected to provide
a scale of supersymmetry (SUSY) breaking
which eventually induces electroweak symmetry breaking
at the weak scale.
Although the simplest possibility
is given by a pure Yang-Mills (YM) theory
in the hidden sector (with singlets)\rlap,
\refmark{\Nil}
it suffers from a cosmological problem
due to overproduction of light particles
called the Polonyi field\rlap,
\refmark{\Cou}
which is a gauge singlet
\foot{An elementary singlet is adequit,
in particular, to produce sizable gaugino and Higgsino masses
(see section 5).}
to gravitationally transmit the SUSY breaking
in the hidden sector to the visible sector.

In this paper, we explore the next simplest candidate
-- vector-like nonabelian gauge theory --
to break SUSY dynamically.
Namely, matter fields in a real representation
are introduced in the hidden sector.
It turns out that this approach circumvents
the Polonyi problem to give viable models for
dynamical SUSY breaking.

%
{\caps 2. QCD-like Interaction in the Hidden Sector}

Let us consider supersymmetric SU($N_c$) gauge theory with
$N_f$ pairs of
hidden quark chiral superfields $Q$ and ${\overline Q}$
in the fundamental representations
$N_c$ and ${\overline N}_c$, respectively.
The hidden quarks are put to be massless\rlap.
\foot{More presicely, the hidden quark mass term is excluded,
which can be imposed
by an axial symmetry given below.}
Otherwise, they would naturally have masses of order
the gravitational scale $M \sim 10^{18}$GeV,
which would make them decouple essentially to yield
an effective theory similar to the pure YM case.

We also introduce a singlet chiral superfield
$Z$, whose lowest component is denoted by $z$.
In contrast to the pure YM case, our QCD-like hidden
sector contains a marginal interaction with the Polonyi
superfield $Z$:
$$
  W_0 = \l_0 ZQ{\overline Q},
 \eqn\INT
$$
which is crucial to avoid the Polonyi problem\rlap.
\foot{Only $M$-suppressed interactions with $Z$ are present
in the pure YM case,
which inevitablly causes the Polonyi problem\rlap.
\refmark{\Ban}}
Here the self interactions of the field $Z$ are excluded
in accord with an axial U$(1)_A$ symmetry
$$
  Z \rightarrow e^{-2i\x}Z, \quad Q \rightarrow e^{i\x}Q,
  \quad {\overline Q} \rightarrow e^{i\x}{\overline Q}
 \eqn\AS
$$
which removes the hidden quark mass term.

It is important that the global symmetry U$(1)_A$ has
a hidden QCD anomaly, which allows a nonperturbative generation
of an effective superpotential for $Z$ violating U$(1)_A$.
In fact,
under the circumsatances that $\VEV{z} \neq 0$,
the hidden gauge and quark superfields may be integrated out
to yield an effective superpotential
\refmark{\Aff}
\foot{$N_f < 3N_c$ for asymptotic freedom.
The form of this effective superpotential is understood
by means of the non-anomolous U$(1)_R$ symmetry
existing in the massless QCD with the superpotential \INT.}
$$
  W_1 = \l \L^{3-n} Z^n; \quad n = {N_f \o N_c},
 \eqn\SP
$$
where $\L$ denotes the hidden
QCD scale, which is supposed to satisfy $|\VEV{z}| \ll \L \ll M$.
On the other hand,
the K{\" a}hler potential then takes the following form:
$$
  K = ZZ^* - {k \o 2\L^2}(ZZ^*)^2 + \cdots,
 \eqn\KP
$$
where $k$ is real. Here the ellipsis denotes higher-order terms of
$ZZ^*$, which are negligible in the subsequent analysis
as far as $|\VEV{z}| \ll \L$.
The dimensionless constants $\l$ and $k$ are
expected to be of order one.
Note that the higher-dimensional terms in Eq.~\KP\ are suppressed
by $\L$ rather than $M$ in contrast to the pure YM case.

We also introduce
a constant term $w$ to get
$$
  W = w + W_1 = w + \l \L^{3-n} Z^n
 \eqn\TEP
$$
as a total effective superpotential,
where $w$ and $\l$ are chosen to be real.
Though such a constant term may be
induced by some dynamics outside the present hidden QCD,
we simply regard it as an input parameter in this paper.

In the following discussion, we restrict
ourselves to the case of $N_f = N_c$,
in which the condition $0 \neq |\VEV{z}| \ll \L$
turns out to be consistent
with the effective potential for $z$ obtained in the next section.

%
{\caps 3. Effective Potential for the Polonyi Field}

The effective potential for the Polonyi field $z$
in the presence of supergravity is given by
$$
  V = \exp\!\left( {K \o M^2} \right)
      \left\{ \left( {\q^2 K \o \q z \q z^*} \right)^{-1}
      \left| {\q W \o \q z} + {\q K \o \q z}{W \o M^2} \right|^2
      - {3|W|^2 \o M^2} \right\}.
 \eqn\EP
$$
Requirement of vanishing cosmological constant on the vacuum
with $|\VEV{z}| \ll \L \ll M$ implies that $|w|M^{-1} < \infty$
when $M \rightarrow \infty$.
Then we obtain an approximate expression for the potential
$$
 \eqalign{
  V &\simeq \left( 1 + {2k \o \L^2}zz^* \right)
     \left| \l \L^2 + z^* {w \o M^2} \right|^2
     - {3 \o M^2}|w + \l \L^2 z|^2 \cr
    &\simeq \l^2 \L^4 - {3 \o M^2} w^2 - {4 \o M^2}w \l \L^2 x
     + 2k \l^2 \L^2 (x^2 + y^2); \quad z = x + iy, \cr
 }
 \eqn\APPR
$$
where $x$ and $y$ are real and we suppose $k > 0$
from convexity of the effective potential
in the limit $M \rightarrow \infty$.
This results in the flat vacuum
\foot{Although it may not correspond to an absolute minimum
of the potential,
this vacuum is practically stable once a sufficiently large
flat universe is formed\rlap.
\refmark{\Wei}}
with
$$
  w \simeq {1 \o \sqrt{3}} \l \L^2 M, \quad
  \VEV{z} \simeq {\L^2 \o \sqrt{3} k M},
 \eqn\VAC
$$
which confirms the consistency $0 \neq |\VEV{z}| \ll \L$.
This vacuum breaks SUSY with the breaking scale
$M_S^2 \simeq \sqrt{3} |\VEV{W}| M^{-1} \sim \L^2$.

%
{\caps 4. Low-energy Contents in the Hidden Sector}

The gravitino mass is given by
$$
  m_{3/2} \simeq {|\VEV{W}| \o M^2} \simeq {\l \L^2 \o \sqrt{3} M},
 \eqn\GM
$$
which characterizes the effective
SUSY breaking scale in the visible sector
and eventually the weak scale.
Hence we set $\L \sim 10^{10}$GeV to get $m_{3/2} \sim 10^2$GeV.

The mass of the Polonyi field is of order $\L$ as seen in Eq.~\APPR,
which shows that the present model does not suffer from
the Polonyi problem as desired.

The interaction \INT\ indicates
that the hidden squarks acquire mass
given by the $F$-component of $Z$, which is of order $M_S^2 \sim \L^2$.
Therefore the low-energy contents in the hidden sector
are effectively described by
non-supersymmetric QCD of hidden quarks
(and `gauginos' -- see section 5)
with mass
$m = \l_0 |\VEV{z}| \sim 10^2$GeV.
Thus there seems to exist hidden pions
(and `$R$-axion'
\refmark{\Bag})
with mass of order $\sqrt{m \L} \sim 10^6$GeV,
which may dominantly decay into gravitinos.
This decay does not produce any cosmological problems
if the reheating temperature after inflation
is sufficiently low\rlap.
\refmark{\Joi}

%
{\caps 5. Gaugino Mass and Quasi-symmetry}

In order to give sizable masses to gauginos such as the gluino,
we introduce higher-dimensional terms
\refmark{\Cre}
of the form
$$
  {1 \o M}ZW_\a W^\a,
 \eqn\GM
$$
where $W_\a$ denote field-strength chiral superfields
for gauge multiplets.
The order of gaugino masses
is then given by $M^{-1} M_S^2 \sim 10^2$GeV\rlap.
\foot{The Higgsino mass is also induced by a higher-dimensional term
\refmark{\Giu}
of the form $M^{-1} Z^* H {\overline H}$
in the K{\" a}hler potential,
where $H$ and ${\overline H}$ denote
doublet Higgs chiral superfields.
Its order is given by $M^{-1} M_S^2$.}

Though higher-dimensional terms are common in supergravity,
the terms \GM\ pose a naturalness problem:
the symmetry U$(1)_A$ defined by \AS\ removes
the hidden quark mass term naturally, whereas
it cannot help excluding \GM\ simultaneously.

Necessity of the terms \GM\ leads us to consider U$(1)_A$ as
a quasi-symmetry -- symmetry broken explicitly only through
gravitational effects.
The quasi-symmetry is supposed to be exact
in the limit $M \rightarrow \infty$.
Thus the presence of $M$-suppressed terms \GM\ is
consistent with the notion of quasi-symmetry\rlap.
\foot{The quasi-symmetry U$(1)_A$ allows nonrenormalizable
interactions such as $M^{-1} Z^4$ in the superpotential,
whose presence never affects our conclusion.}
Moreover, it may be argued that a global symmetry is
naturally a quasi-symmetry since its global charge `dissipates'
through black holes or wormholes
\refmark{\Gid}
in quantum theory of gravitation\rlap.
\foot{However, it is not clear to us whether breaking terms
of the global symmetry are always suppressed by the gravitational
scale $M$.}

%
{\caps 6. Conclusion}

We have considered QCD-like hidden sector models
with the superpotential \INT\ on which
the quasi-symmetry \AS\ is implimented.
They turned out to suffer from no cosmological problems
and to give sizable gaugino masses.
Thus we conclude that they
constitute simple and viable models
of SUSY breaking in the hidden sector.

%
\acknowledge

We would like to thank S.~Tanimura
for valuable discussions.
I.~K.-I.~is grateful to H.~Nakano for stimulating discussions.


\endpage

\refout

\bye